\documentclass[%
 aip,
 amsmath,amssymb,
 reprint,%
]{revtex4-1}

\usepackage{graphicx}
\graphicspath{{figures/}}

\usepackage{subcaption}
\usepackage{dcolumn}
\usepackage{bm}

\usepackage[utf8]{inputenc}
\usepackage[T1]{fontenc}
\usepackage{mathptmx}
\usepackage{xcolor}

\newcommand{\blue}[1]{{\color{black}{#1}}}
\begin{document}

\preprint{AIP/123-QED}

\title{Visualizing droplet dispersal for face shields and masks with exhalation valves}

\author{Siddhartha Verma}
\email{vermas@fau.edu}
\homepage{http://www.computation.fau.edu}
\altaffiliation[Also at ]{Harbor Branch Oceanographic Institute, Florida Atlantic University, Fort Pierce, FL 34946, USA}
\author{Manhar Dhanak}%
\email{dhanak@fau.edu}
\author{John Frankenfield}
\email{jfranken@fau.edu}
\affiliation{Department of Ocean and Mechanical Engineering, Florida Atlantic University, Boca Raton,\\
FL 33431, USA
}%
\date{\today}

\begin{abstract}
Several places across the world are experiencing a steep surge in COVID-19 infections. Face masks have become increasingly accepted as one of the most effective means for combating the spread of the disease, when used in combination with social-distancing and frequent hand-washing. However, there is an increasing trend of people substituting regular cloth or surgical masks with clear plastic face shields, and with masks equipped with exhalation valves. One of the factors driving this increased adoption is improved comfort compared to regular masks. However, there is a possibility that widespread public use of these alternatives to regular masks could have an adverse effect on mitigation efforts. To help increase public awareness regarding the effectiveness of these alternative options, we use qualitative visualizations to examine the performance of face shields and exhalation valves in impeding the spread of aerosol-sized droplets. The visualizations indicate that although face shields block the initial forward motion of the jet, the expelled droplets can move around the visor with relative ease and spread out over a large area depending on \blue{light ambient disturbances}. Visualizations for a mask equipped with an exhalation port indicate that a large number of droplets pass through the exhale valve unfiltered, which significantly reduces its effectiveness as a means of source control. Our observations suggest that to minimize the community spread of COVID-19, it may be preferable to use high quality cloth or surgical masks that are of a plain design, instead of face shields and masks equipped with exhale valves.
\end{abstract}

\maketitle

The COVID-19 pandemic has deeply affected every aspect of daily life worldwide. Several places across the world, including the United States, Brazil, and India, are experiencing a steep surge in infections. Healthcare systems in the most severely affected locations have been stretched to capacity, which also tends to impact urgent care for cases unrelated to COVID-19~\cite{Emanuel2020,Soreide2020}. Researchers have reported steady progress in the development of potential treatments and vaccines, however, it is estimated that widespread inoculation will not be available until sometime in the year 2021. It appears that the likelihood of vulnerable individuals struggling with severe health issues, and debilitating socio-economic ramifications of the pandemic, will continue in the foreseeable future. Furthermore, widespread uncertainty regarding the re-opening of schools and universities for in-person instruction has created additional cause for concern, since these institutions have the potential to become focal points for unchecked community spread of the disease. In light of the acute circumstances, it has become crucial to establish clear and specific guidelines that can help mitigate the disease's spread, especially given the high prevalence of asymptomatic and pre-symptomatic spread~\cite{CentersforDiseaseControlandPrevention2020}. \blue{A number of recent studies have contributed to this effort, by significantly improving our understanding of various physical mechanisms involved in the disease's transmission~\cite{Dbouk2020,Bhardwaj2020,Busco2020,Chaudhuri2020}.}

Face masks have become increasingly accepted as one of the most effective means for source control (i.e., protecting others from a potentially infected wearer), and can help curb the community spread of the disease when used in combination with social-distancing and frequent hand-washing~\cite{Cheng2020,Clase2020,Brooks2020,CentersforDiseaseControlandPrevention2020a,Dbouk2020Masks}. Widespread \blue{mask-use} by the general population has now been recommended or mandated in various places and communities around the world. Several private businesses have also adopted requirements for customers to use face coverings. Importantly, certain cloth-based masks have been shown to be effective in blocking the forward spread of aerosolized droplets~\cite{Verma2020_PoF} (Supplementary Movie 1). Although they are somewhat less capable than well-fitted \blue{medical grade} masks, homemade masks constructed using certain materials can filter out a large proportion of respiratory droplets and particles~\cite{Rengasamy2010,Davies2013,Bae2020,Konda2020, Lustig2020, Ma2020}. Moreover, cloth masks have the advantage of being readily available to the wider public, in addition to being cost-effective, comfortable, and reusable when washed and disinfected properly. Additionally, they do not divert away from the supply of \blue{medical grade} masks for healthcare workers.

\begin{figure}[ht!]
\centering
\begin{subfigure}{0.49\linewidth}
\centering
\includegraphics[width=\linewidth]{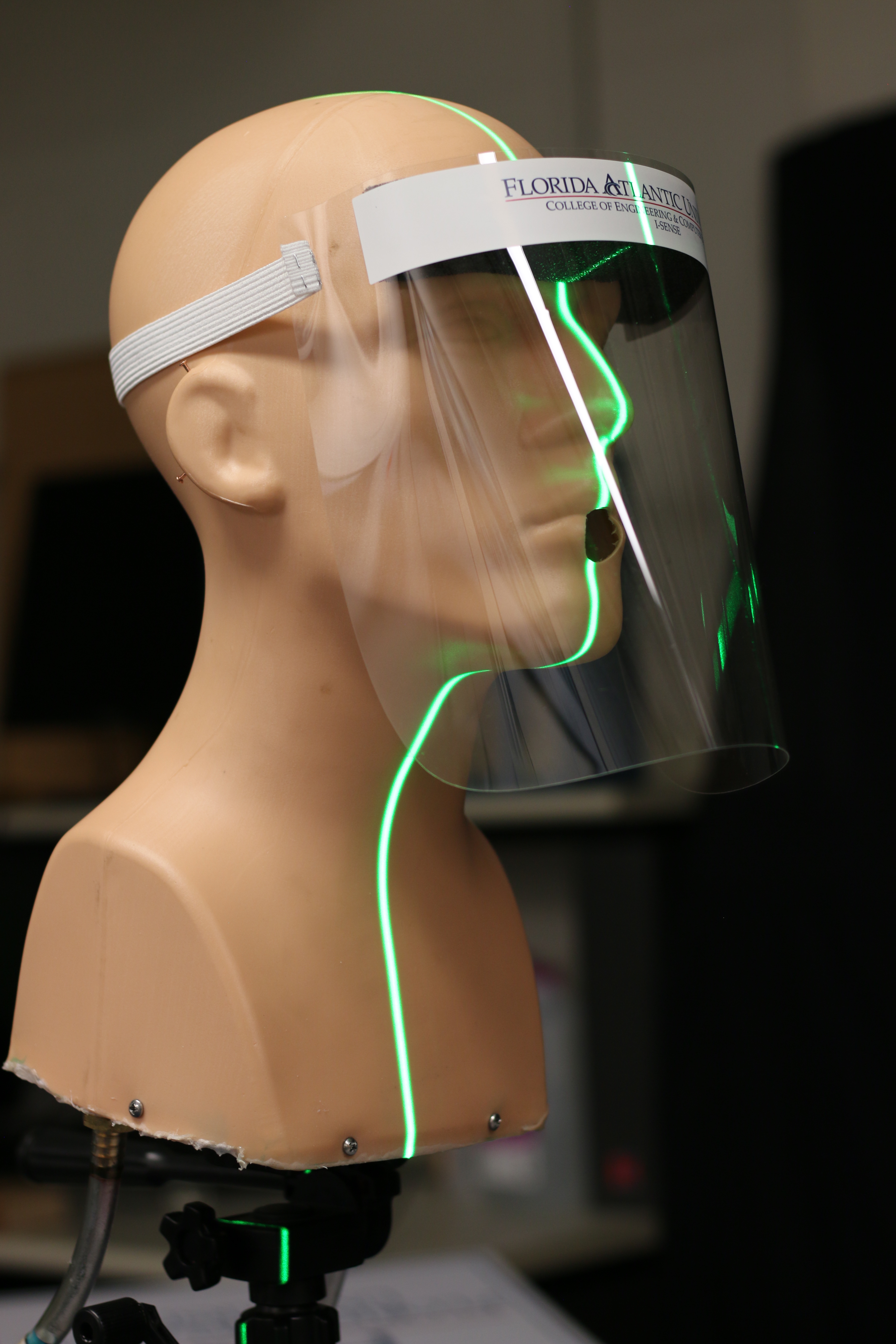}
\caption{\label{fig:dayShield}}
\end{subfigure}
\begin{subfigure}{0.49\linewidth}
\centering
\includegraphics[width=\linewidth]{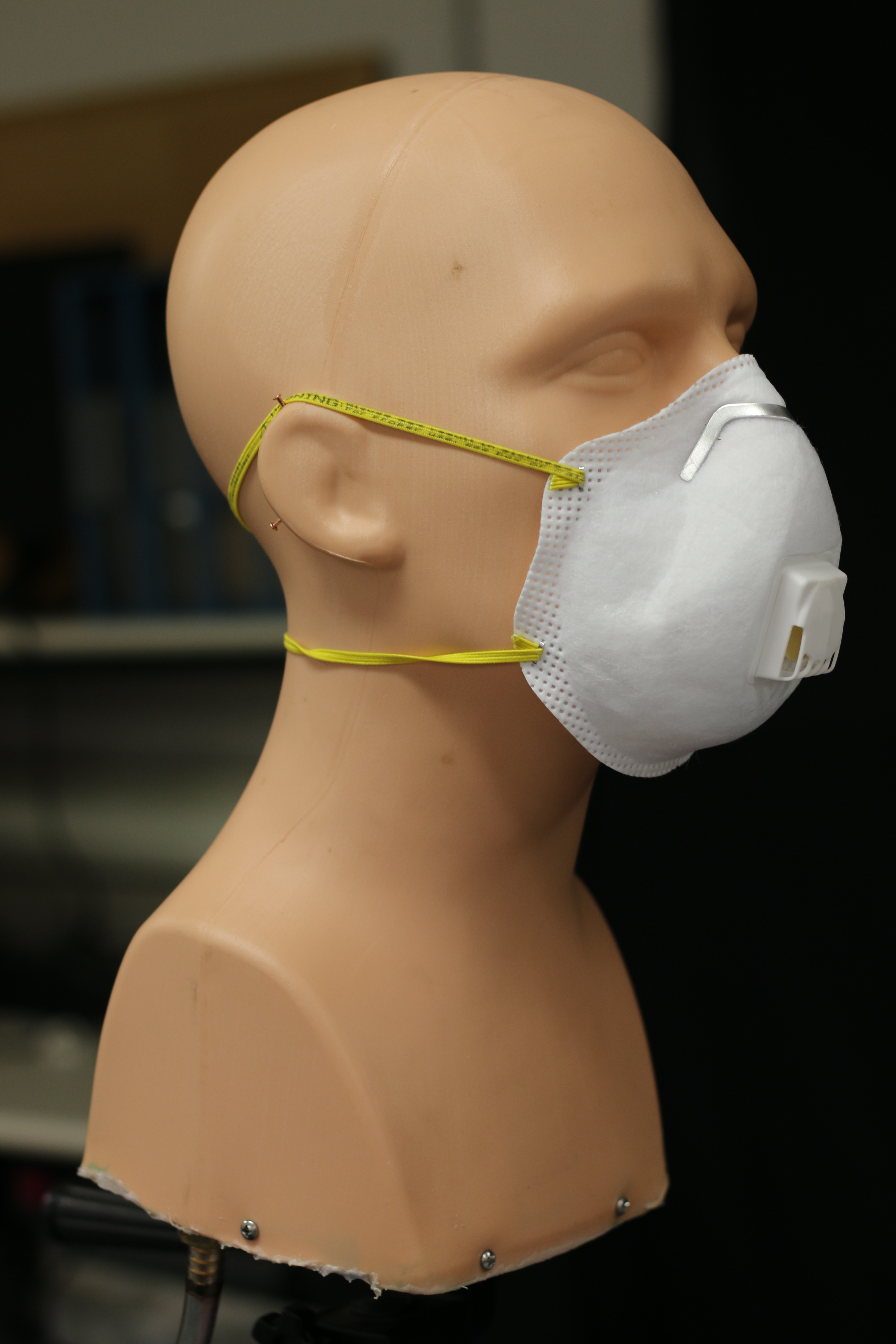}
\caption{\label{fig:dayValve}}
\end{subfigure}
\caption{\label{fig:daylight}(\subref{fig:dayShield}) A face shield, which is similar in design to those used by healthcare workers in conjunction with masks and other protective equipment. The vertical laser sheet used for visualizing the expelled droplets is visible in this panel. (\subref{fig:dayValve}) An N95 mask with an exhalation valve located at the front. Both cloth-based and N95 masks can be found equipped with such exhalation ports.}
\end{figure}

While broad acceptance regarding the need for face coverings has risen steadily, there is an increasing trend of people substituting regular cloth or surgical masks with clear plastic face shields, and with masks equipped with exhalation valves (Figure~\ref{fig:daylight}).
Face shields tend to have noticeable gaps along the bottom and the sides, and are used in the medical community primarily for protecting the wearer against incoming sprays and splashes while in close proximity to patients~\cite{Roberge2016}. Moreover, they tend to be used in conjunction with respirators, surgical masks, or other protective equipment. Masks with exhalation ports include a one-way valve which restricts airflow when breathing in, but allows free outflow of air. The inhaled air gets filtered through the mask material, however, the exhaled breath passes through the valve unfiltered. There has been \blue{limited} research on how effective face shields \blue{and masks with exhalation valves are as a means of source control~\cite{Ippolito2020,Viola2020}}.

One of the factors driving the increased adoption of shields and exhalation valves is improved comfort compared to regular surgical or cloth masks. Exhale valves allow for improved breathability, and reduce humidity and fogging \blue{when wearing prescription glasses}. Face shields also offer these benefits, in addition to protecting the eyes from splashes and sprays of infected droplets~\cite{Lindsley2014,Roberge2016}. Shields have also been credited with other advantages such as ease of cleaning and disinfecting, long-term reusability (which is also true for well-constructed cloth masks), and allowing visual communication of facial expressions for people who may be hearing-impaired~\cite{Roberge2016,Perencevich2020}. 

Notably, a recent opinion-based article by Perencevich et al.~\cite{Perencevich2020} suggested that shields may be a better alternative to regular masks for combating the COVID-19 crisis. The authors' opinion is based on the premise that ejecta from the mouth and nose hit the visor, and their forward motion is arrested completely. While this is true for relatively large respiratory droplets, the effect on the smaller aerosol-sized droplets (diameter less than approximately $5\mu m-10 \mu m$) is markedly different, since they act as tracers and have a higher tendency to follow airflow patterns more faithfully. We note that one of the primary studies cited by Perencevich et al. expressly indicates that face shields did not serve as primary respiratory protection for the wearer in experimental tests, since suspended aerosols could flow around the visor and enter the respiratory tract~\cite{Lindsley2014}. Over an exposure duration of 1 to 30 minutes, the shield was only $23\%$ effective in reducing inhalation of droplets that were $3.4\mu m$ on average. Although this study by Lidsley et al.~\cite{Lindsley2014} did not consider face shields as source control methods, they are likely to suffer the same disadvantage in this role, since smaller outgoing droplets will flow around the bottom and the sides of the visor. While the opinion article by Perencevich et al.~\cite{Perencevich2020} is based on the presumption that transmission of COVID-19 occurs primarily through larger respiratory droplets, recent studies support the possibility of airborne transmission via aerosol-sized droplets~\cite{Morawska2020,Liu2020,Ong2020,Cai2020}. 

Current CDC guidelines discourage the use of face shields as a sole means of source control~\cite{CentersforDiseaseControlandPrevention2020_Cloth}, and mention that masks equipped with exhalation valves should not be used when a sterile environment is required~\cite{CentersforDiseaseControlandPrevention2020_Valve}. At the same time, there are broad variations in recommendations made by states and counties across the US, with some allowing the use of face shields as alternatives to masks~\cite{Oregon,NCarolina}, whereas many others do not address the issue at all. There is a possibility that widespread public adoption of these alternatives to regular masks could have an adverse effect on mitigation efforts. To help increase public awareness regarding the effectiveness of these alternative options, we use qualitative visualizations to examine the performance of face shields and exhale valves in impeding droplet spread.

The visualization setup used here is similar to the arrangement used in a prior study~\cite{Verma2020_PoF} which examined the effectiveness of various facemasks in stopping the spread of respiratory jets. The setup consists of a hollow manikin head, where a cough/sneeze is emulated via a pressure impulse applied using a manual pump. \blue{The air capacity of the pump is $500ml$, which is comparable to the lower end of the total volume expelled during a cough~\cite{Gupta2009}}. Tracers composed of droplets of distilled water and glycerin are expelled through the mouth opening, and are visualized using laser sheets to observe the spatial and temporal development of the ejected flow. Up to two laser sheets are used in the visualizations presented here, to provide a better indication of the volumetric spread of the expelled jet. The tracer droplets' diameter was estimated to be less than $10 \mu m$, based on Stokes' law and the observation that they could remained suspended in a quiescent environment for between 2 to 3 minutes \blue{without significant settling}.
\blue{The settling velocity for spheres in Stokes flow (i.e., at very low Reynolds numbers) is given as follows:
\begin{equation}
v = \dfrac{(\rho_p - \rho_f)gd^2}{18 \mu},
\end{equation}
where $\rho_p$ is the density of the spherical particle, $\rho_f$ is the density of the ambient fluid (air), $\mu$ is the dynamic viscosity, $g$ is acceleration due to gravity, and $d$ is the diameter of the sphere. Using the density of water as an approximation for $\rho_p$, and $d=1e-5m$, we get a settling speed of $0.003m/s$. Thus, a droplet of diameter $10 \mu m$ would fall a distance of $0.45m$ in $150s$, i.e., in 2.5 minutes. From our observations in a minimally disturbed environment, the droplets did not display significant downward gravity-driven settling within this time-frame. The droplets eventually disappeared from view, either because they moved laterally off the laser sheet, or because they experienced further evaporation.
}
Additional details regarding the visualization setup may be found in ref.~\cite{Verma2020_PoF}. \blue{We remark that the all of the flows described in this work are inherently three-dimensional in nature, but the visualizations only depict plane two-dimensional cross-sections. For instance, the uncovered emulated cough shown in Supplementary Movie 1 displays three-dimensional behavior, such as the motion of the leading plume which resembles the formation of a vortex ring. The lateral (sideways) motion of the jets is also evident at times when the visible droplet patches disappear or re-appear in the laser sheet.}

\begin{figure*}[ht!]
\centering
\begin{subfigure}{0.49\linewidth}
\centering
\includegraphics[width=\linewidth]{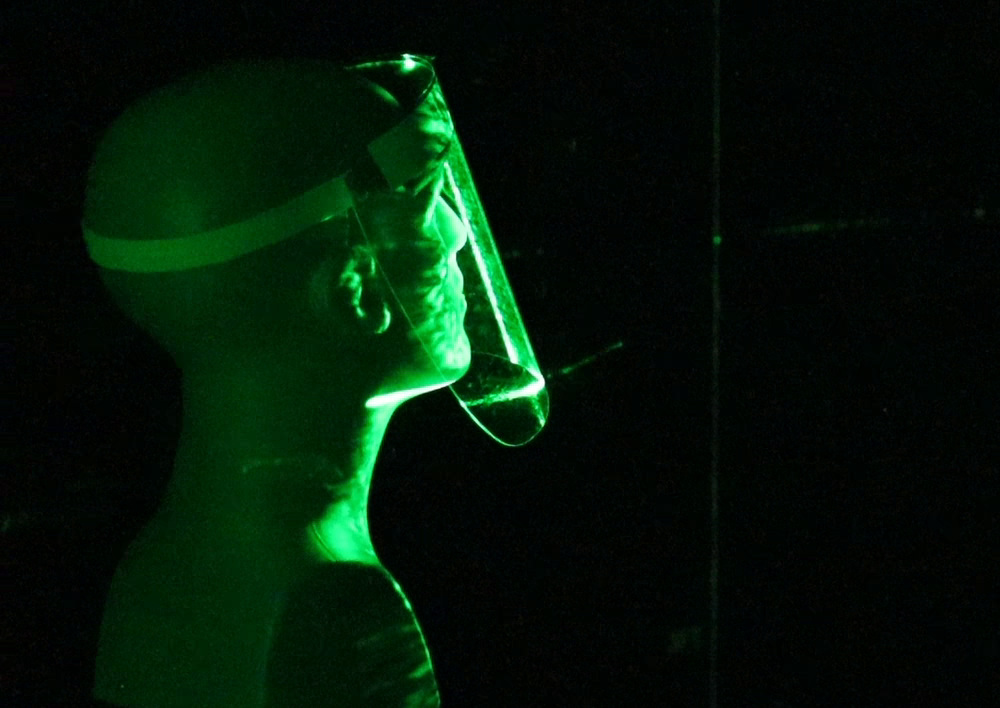}
\caption{\label{fig:shield1}}
\end{subfigure}
\begin{subfigure}{0.49\linewidth}
\centering
\includegraphics[width=\linewidth]{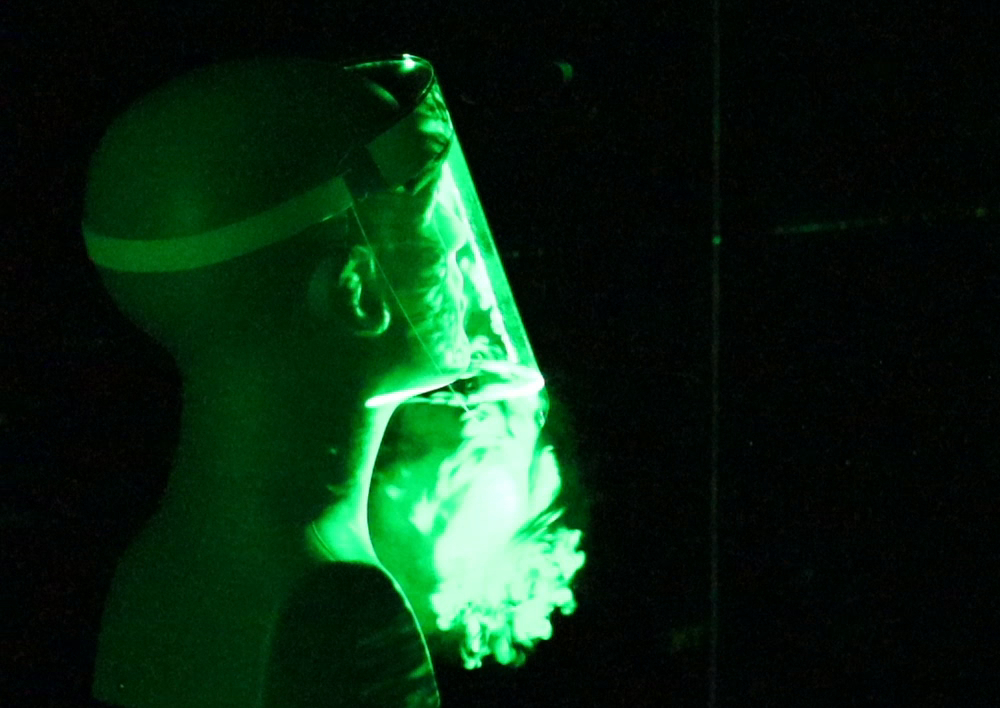}
\caption{\label{fig:shield2}}
\end{subfigure}
\begin{subfigure}{0.49\linewidth}
\centering
\includegraphics[width=\linewidth]{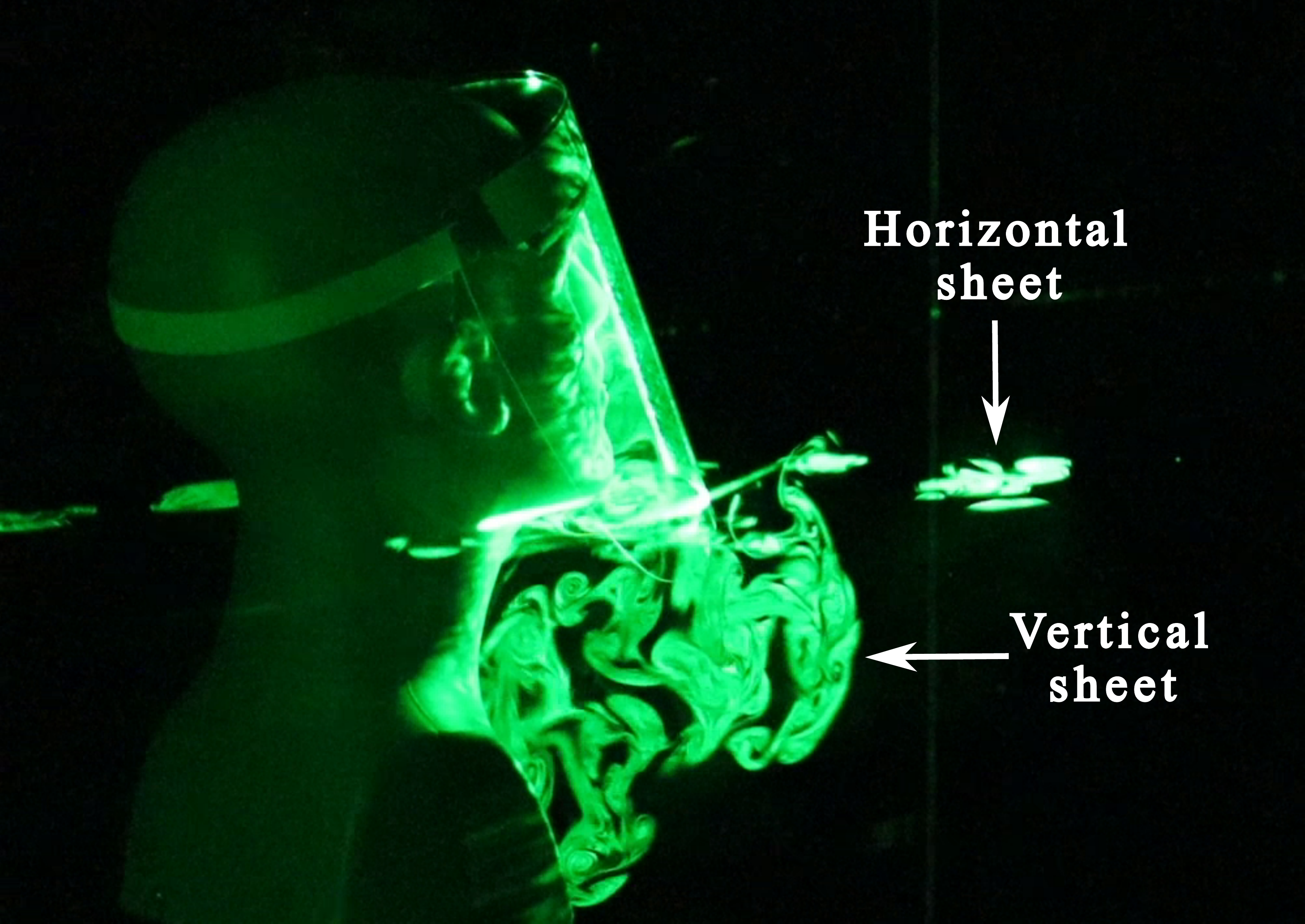}
\caption{\label{fig:shield3}}
\end{subfigure}
\begin{subfigure}{0.49\linewidth}
\centering
\includegraphics[width=\linewidth]{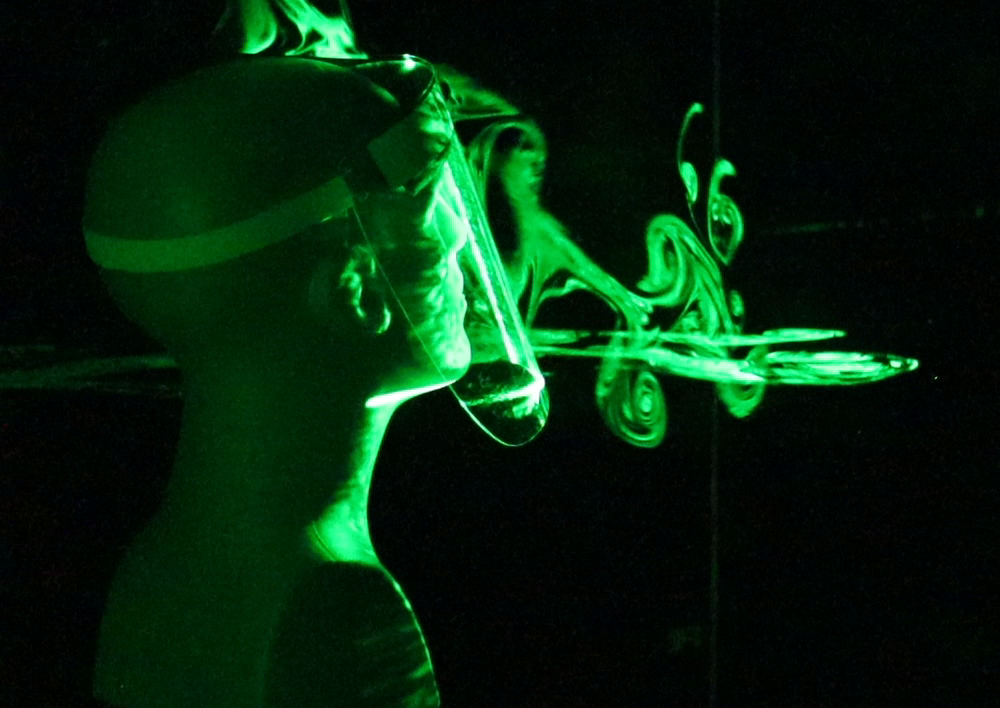}
\caption{\label{fig:shield4}}
\end{subfigure}
\caption{\label{fig:shield} Near-field view of droplet spread when a face shield is used to impede the emerging jet. (\subref{fig:shield1}) Prior to emulating a cough/sneeze. (\subref{fig:shield2}) 0.57 seconds after \blue{the} initiation of the emulated cough. (\subref{fig:shield3}) 3.83 seconds. (\subref{fig:shield4}) 16.57 seconds. The ejected plume is illuminated by both a vertical and a horizontal laser sheet. Droplets illuminated by the horizontal laser sheet can be observed in (\subref{fig:shield3}) and (\subref{fig:shield4}). (Multimedia View)}
\end{figure*}

Figure~\ref{fig:shield} (Multimedia View) shows the evolution of a cough/sneeze when a face shield is used to impede the expelled jet.
As expected, the visor initially deflects the expelled droplets downward (Figure~\ref{fig:shield2}). However, the aerosol-sized droplets do not fall to the ground, but stay suspended beneath the bottom opening of the shield (Figure~\ref{fig:shield3}). These droplets rise upward after a few seconds since they are warmer than the ambient air \blue{owing to the vaporized glycerin-water mixture, and also because they might undergo further evaporation once released into the environment}. A horizontal laser sheet has been used in addition to a vertical sheet, and the lateral spread of the droplets becomes visible as they cross the horizontal plane in Figures~\ref{fig:shield3} and~\ref{fig:shield4}. Although the case depicted here shows droplets spreading in the forward direction, slight variations in ambient \blue{disturbances} were observed to \blue{reverse} the direction of spread\blue{, i.e., towards the manikin's back}. The time evolution of droplet spread from an additional run with a faceshield can be seen in Supplementary Movie 2.

\begin{figure*}[ht!]
\centering
\begin{subfigure}{0.8\linewidth}
\centering
\includegraphics[width=\linewidth]{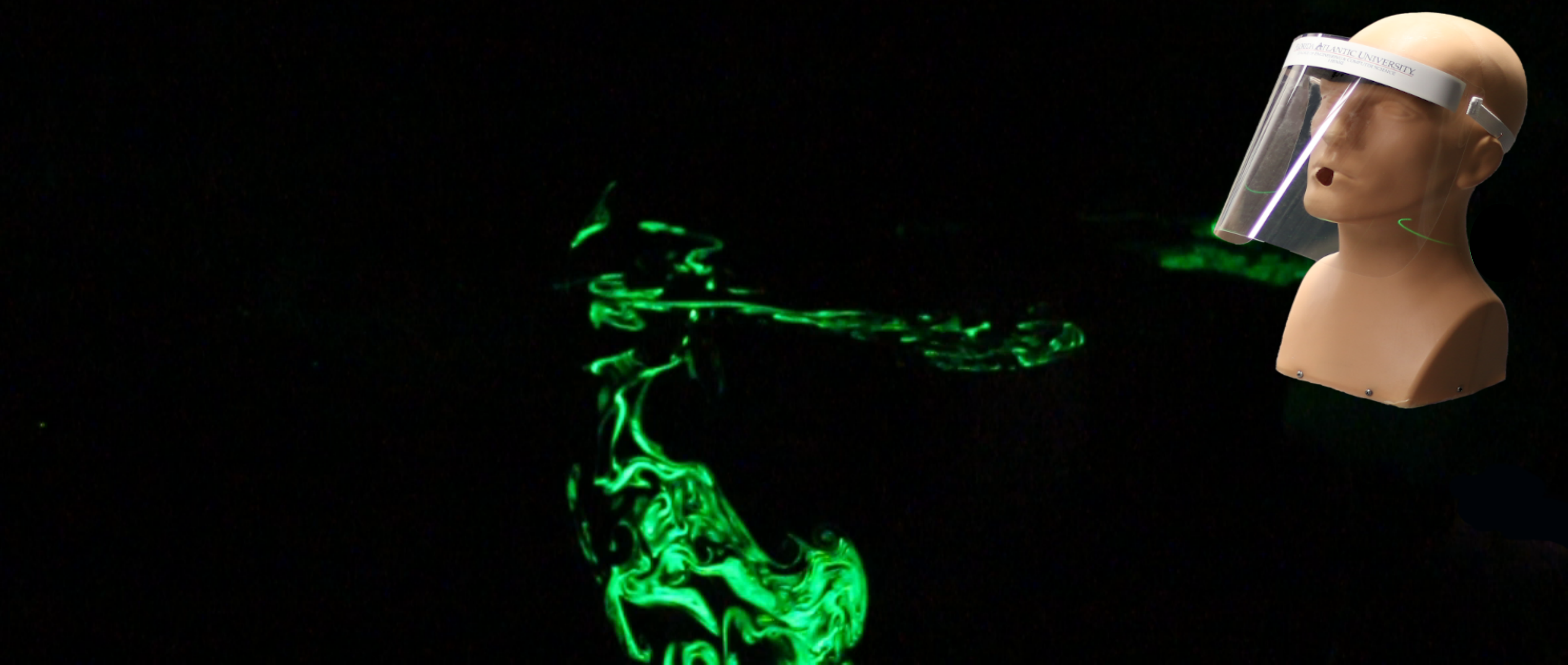}
\caption{\label{fig:farfield1}}
\end{subfigure}
\begin{subfigure}{0.8\linewidth}
\centering
\includegraphics[width=\linewidth]{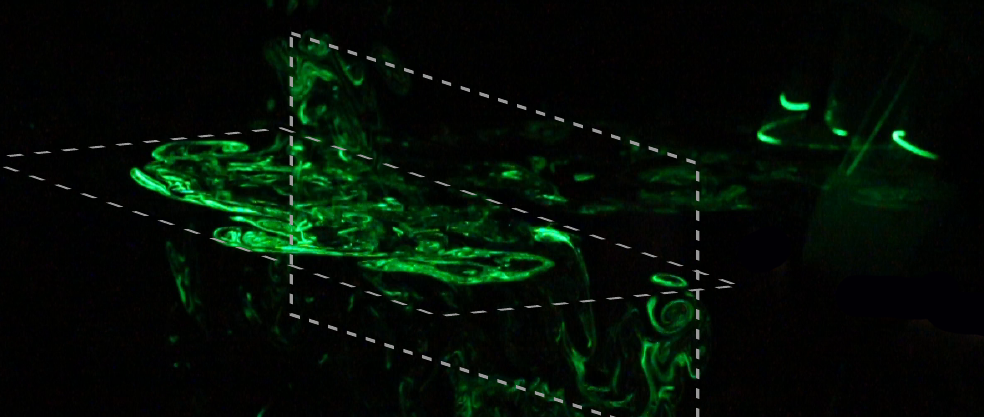}
\caption{\label{fig:farfield2}}
\end{subfigure}
\begin{subfigure}{0.8\linewidth}
\centering
\includegraphics[width=\linewidth]{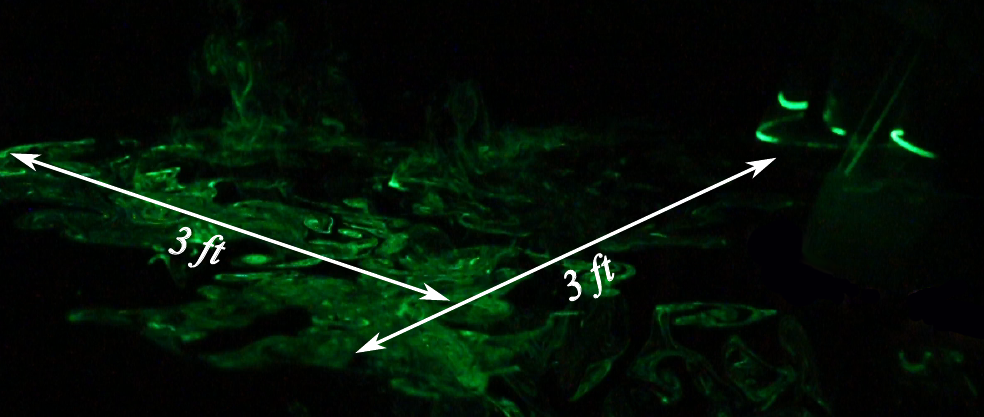}
\caption{\label{fig:farfield3}}
\end{subfigure}
\caption{\label{fig:farfield} Far-field view of droplet spread when a face shield is used to impede the jet. (\subref{fig:farfield1}) 2.97 seconds after \blue{the} initiation of the emulated cough. (\subref{fig:farfield2}) 6.98 seconds. (\subref{fig:farfield3}) 10.77 seconds. (Multimedia view)}
\end{figure*}

To observe the lateral and longitudinal spread of the suspended droplets over a large area, we examine a far-field view in Figure~\ref{fig:farfield} (Multimedia View).
The manikin's position is shown as an overlay in Figure~\ref{fig:farfield1}, where the ejected droplets are visible in a horizontal and a vertical laser sheet, which help convey the spread of the droplets in the lateral and longitudinal directions. The positioning of the laser planes is depicted in Figure~\ref{fig:farfield2}. After 10 seconds (Figure~\ref{fig:farfield3}), the droplets were observed to have spread approximately 3 feet in both the forward and lateral directions. We note that the intensity of the scattered light has decreased noticeably by this point, which is indicative of a decrease in droplet concentration due to spreading over a large volume. \blue{Most of the droplets visible in Figure~\ref{fig:farfield3} are illuminated by the horizontal sheet shown in Figure~\ref{fig:farfield2}, which is aligned with the bottom opening of the face shield.} Very few droplets are visible in the vertical laser sheet, since most of them have advected forward of the sheet's position by this time. \blue{We remark that both the longitudinal and lateral spread depend on a combination of the initial momentum of the cough and advection by very light ambient disturbances. While the specific case discussed here depicts forward spread, we observed that it was equally likely that the droplets could spread in the reverse direction, i.e., behind the manikin. We do not expect diffusion to play a dominant role at the time scales discussed here.} Overall, we can surmise that the face shield blocks the initial forward motion of the jet, however the aerosolized droplets that are expelled can disperse over a wide area over time, albeit with decreasing droplet concentration.

\begin{figure*}[ht!]
\centering
\begin{subfigure}{0.245\linewidth}
\centering
\includegraphics[width=\linewidth]{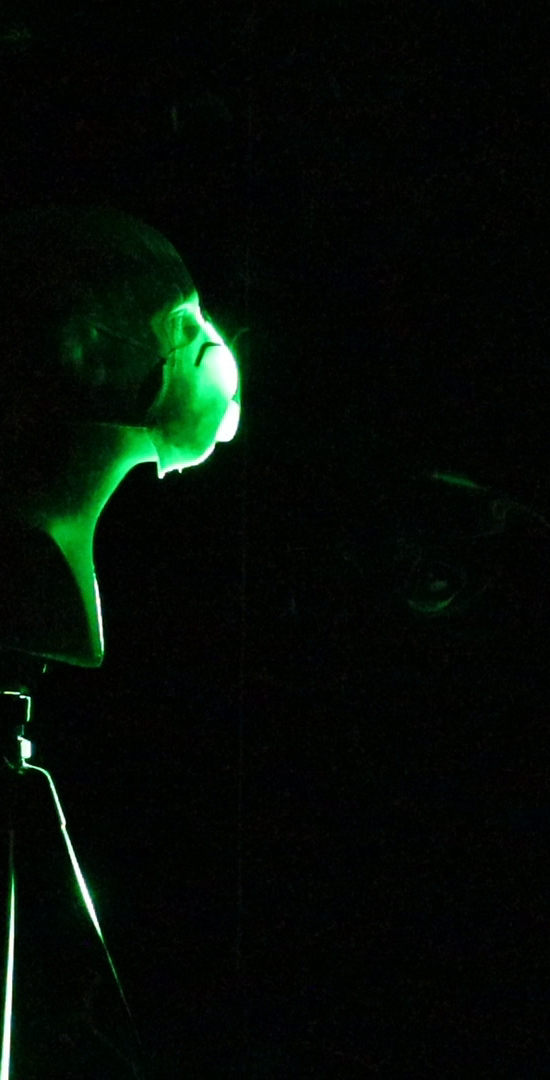}
\caption{\label{fig:valve1}}
\end{subfigure}
\begin{subfigure}{0.245\linewidth}
\centering
\includegraphics[width=\linewidth]{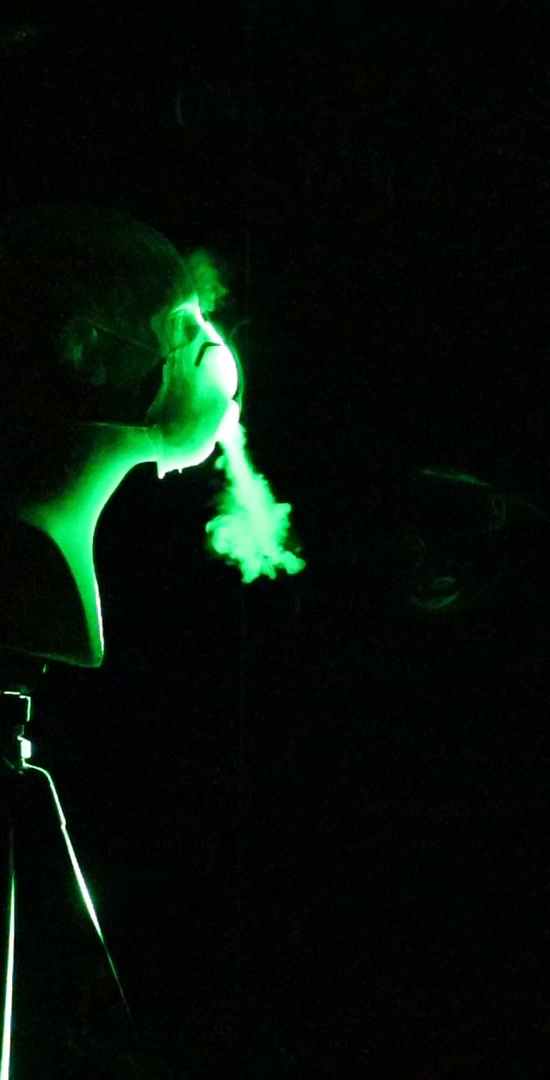}
\caption{\label{fig:valve2}}
\end{subfigure}
\begin{subfigure}{0.245\linewidth}
\centering
\includegraphics[width=\linewidth]{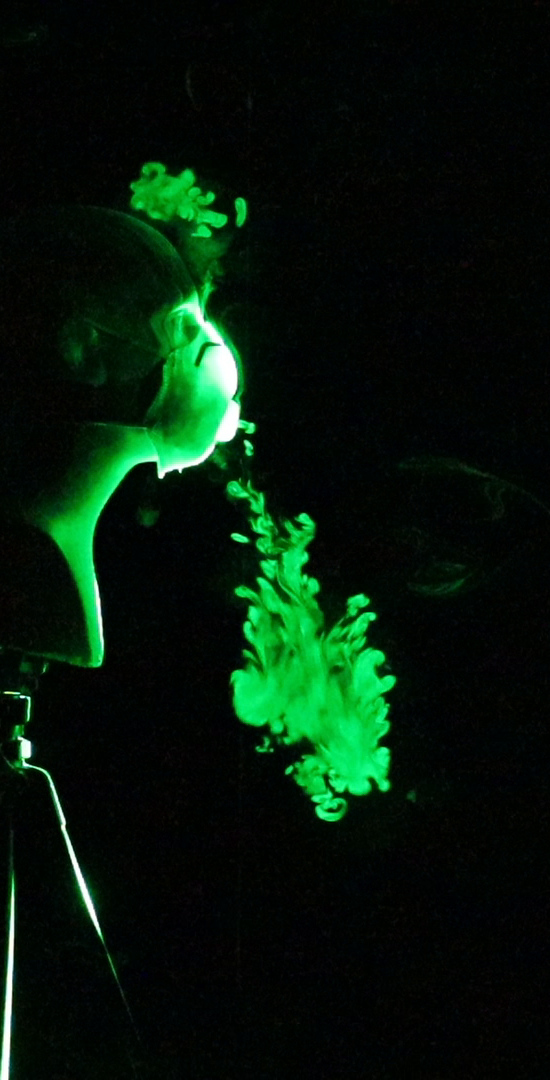}
\caption{\label{fig:valve3}}
\end{subfigure}
\begin{subfigure}{0.245\linewidth}
\centering
\includegraphics[width=\linewidth]{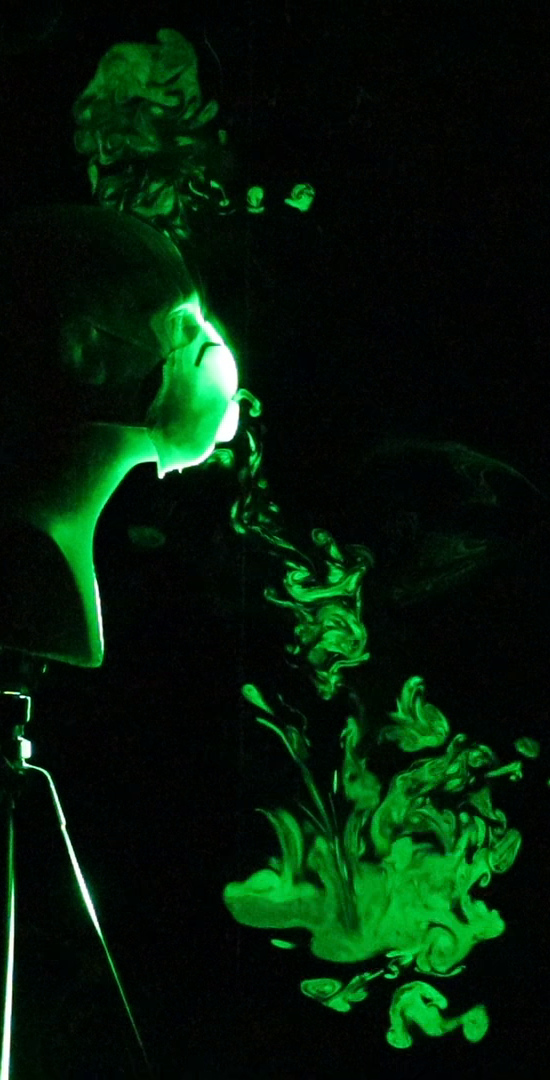}
\caption{\label{fig:valve4}}
\end{subfigure}
\caption{\label{fig:valve} Visualization of droplet \blue{spread} when an N95 mask equipped with an exhalation port is used to impede the emerging jet. (\subref{fig:valve1}) Prior to emulating a cough/sneeze. (\subref{fig:valve2}) 0.2 seconds after \blue{the} initiation of the emulated cough. (\subref{fig:valve3}) 0.63 seconds. (\subref{fig:valve4}) 1.67 seconds. (Multimedia View)}
\end{figure*}

We now consider the effectiveness of masks equipped with exhalation valves in restricting the spread of respiratory droplets. Figure~\ref{fig:valve} (Multimedia View) shows the spatial and temporal evolution of the jets that emerge from an N95 mask which has a single exhalation port located at the front.
Apart from the design used here, certain cloth-based masks that are available commercially also come equipped with one to two exhale ports, located on either side of the facemask. In Figures~\ref{fig:valve2} and~\ref{fig:valve3}, we observe that a small amount of the exhaled droplets escape from the gap between the top of the mask and the bridge of the nose. However, a majority of the exhaled air passes through the exhale port unhindered. The resulting jet is deflected downward in the current case, which reduces the initial forward spread of the droplets. However, the aerosolized droplets will eventually disperse over a large area depending on the ambient \blue{disturbances and} airflow patterns, as in the case of the face shield (Figure~\ref{fig:farfield}). 

\begin{figure*}[ht!]
\centering
\begin{subfigure}{0.245\linewidth}
\centering
\includegraphics[width=\linewidth]{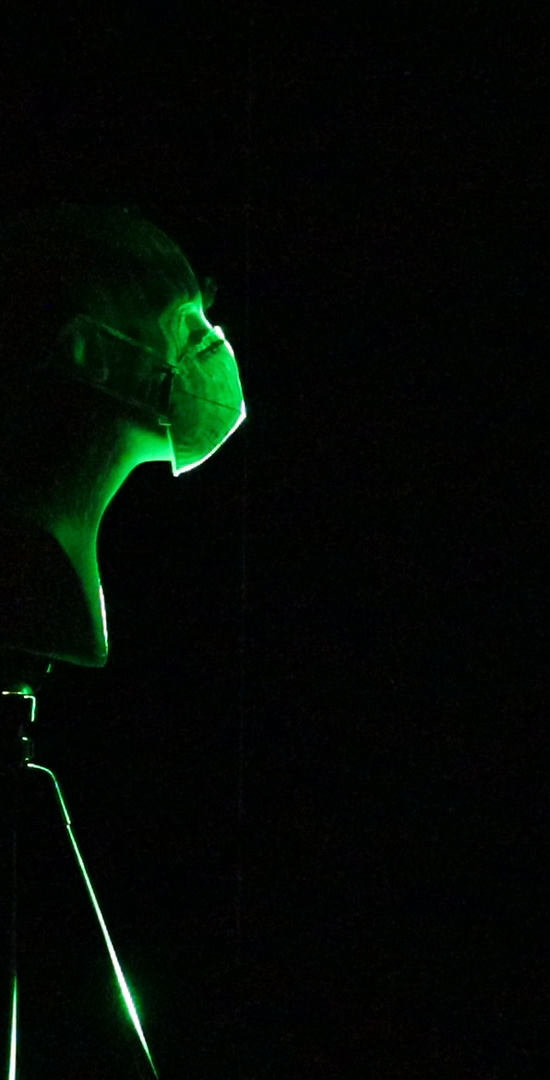}
\caption{\label{fig:n95_1}}
\end{subfigure}
\begin{subfigure}{0.245\linewidth}
\centering
\includegraphics[width=\linewidth]{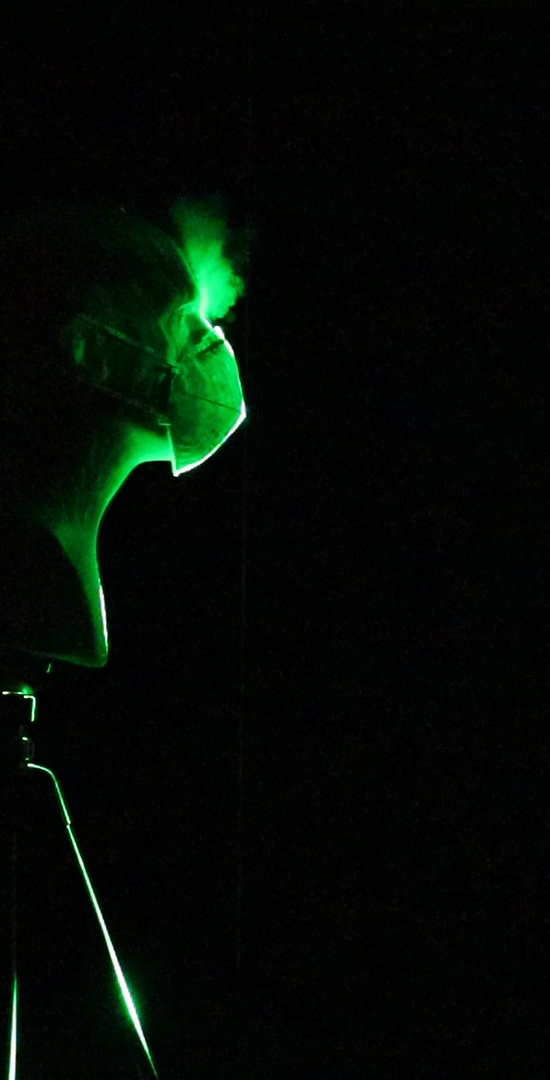}
\caption{\label{fig:n95_2}}
\end{subfigure}
\begin{subfigure}{0.245\linewidth}
\centering
\includegraphics[width=\linewidth]{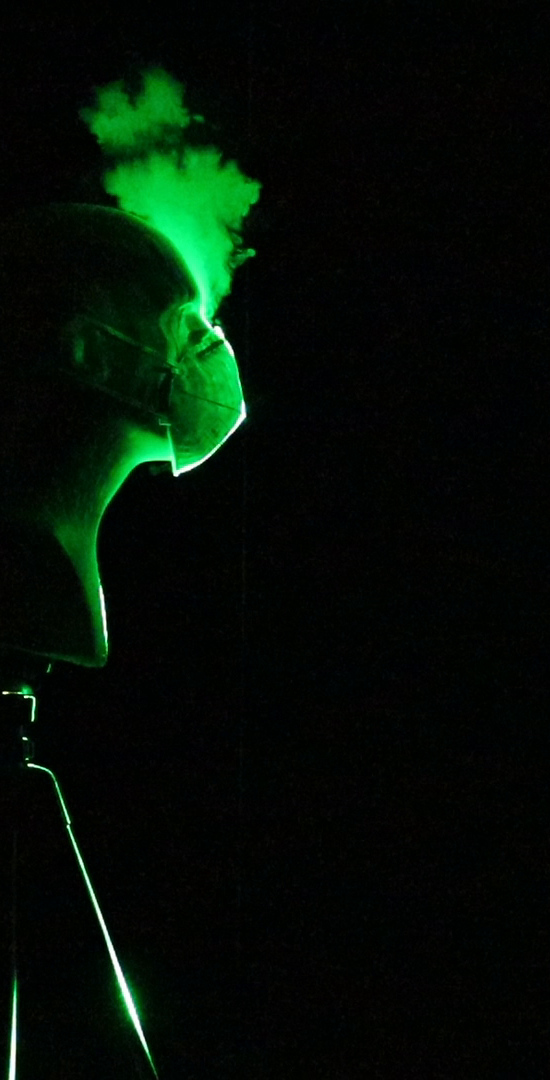}
\caption{\label{fig:n95_3}}
\end{subfigure}
\begin{subfigure}{0.245\linewidth}
\centering
\includegraphics[width=\linewidth]{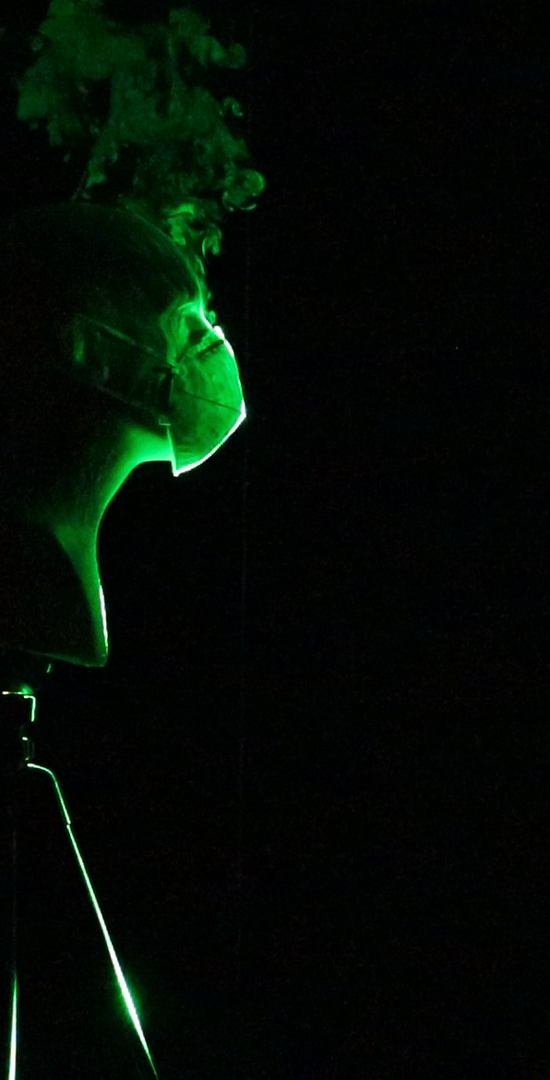}
\caption{\label{fig:n95_4}}
\end{subfigure}
\caption{\label{fig:n95} Visualization of droplet \blue{spread} when a regular N95-rated mask is used to impede the jet. (\subref{fig:n95_1}) Prior to emulating a cough/sneeze. (\subref{fig:n95_2}) 0.13 seconds after \blue{the} initiation of the emulated cough. (\subref{fig:n95_3}) 0.33 seconds. (\subref{fig:n95_4}) 0.83 seconds. (Multimedia View)}
\end{figure*}

We now examine the droplet dispersal pattern when using a regular N95-rated mask in Figure~\ref{fig:n95} (Multimedia View).
Once again, we observe droplets escaping from the gap between the mask and the nose, however, the intensity of light scattered by the escaped droplets is lower than that for the valved N95 mask (Figure~\ref{fig:valve4}). We note that the droplets that escape from the regular mask will also get dispersed by ambient \blue{disturbances}, however, the extent of exposure will be lower compared to that for either face shields or masks with valves. While the two masks shown in Figures~\ref{fig:valve} and~\ref{fig:n95} are N95-rated, we expect the observations described here (with regard to valves) to hold true even for cloth/surgical masks that are of a plain functional design versus those equipped with exhale valves. 

\begin{figure*}[ht!]
\centering
\begin{subfigure}{0.49\linewidth}
\centering
\includegraphics[width=\linewidth]{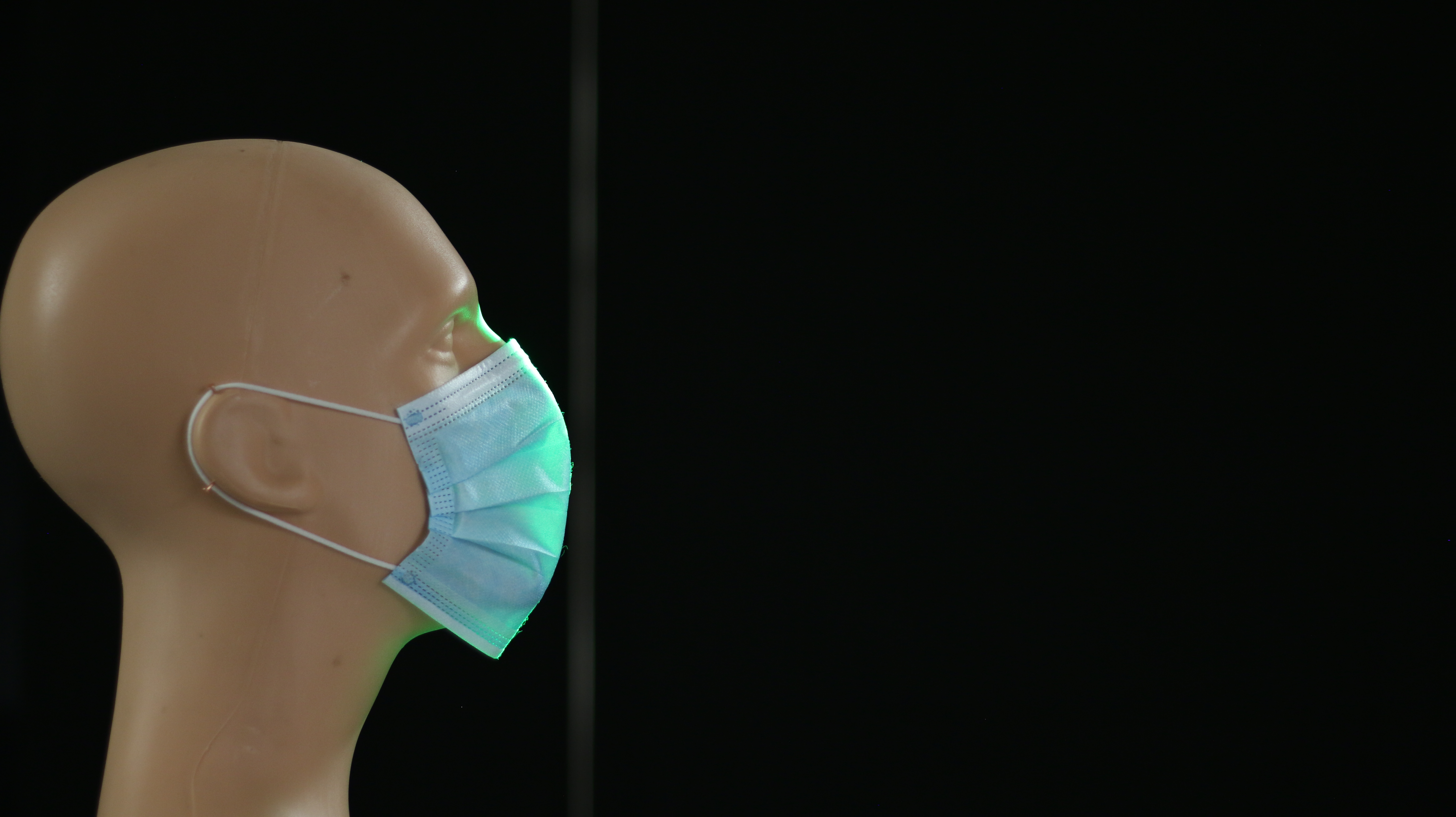}
\caption{\label{fig:surgA1}}
\end{subfigure}
\begin{subfigure}{0.49\linewidth}
\centering
\includegraphics[width=\linewidth]{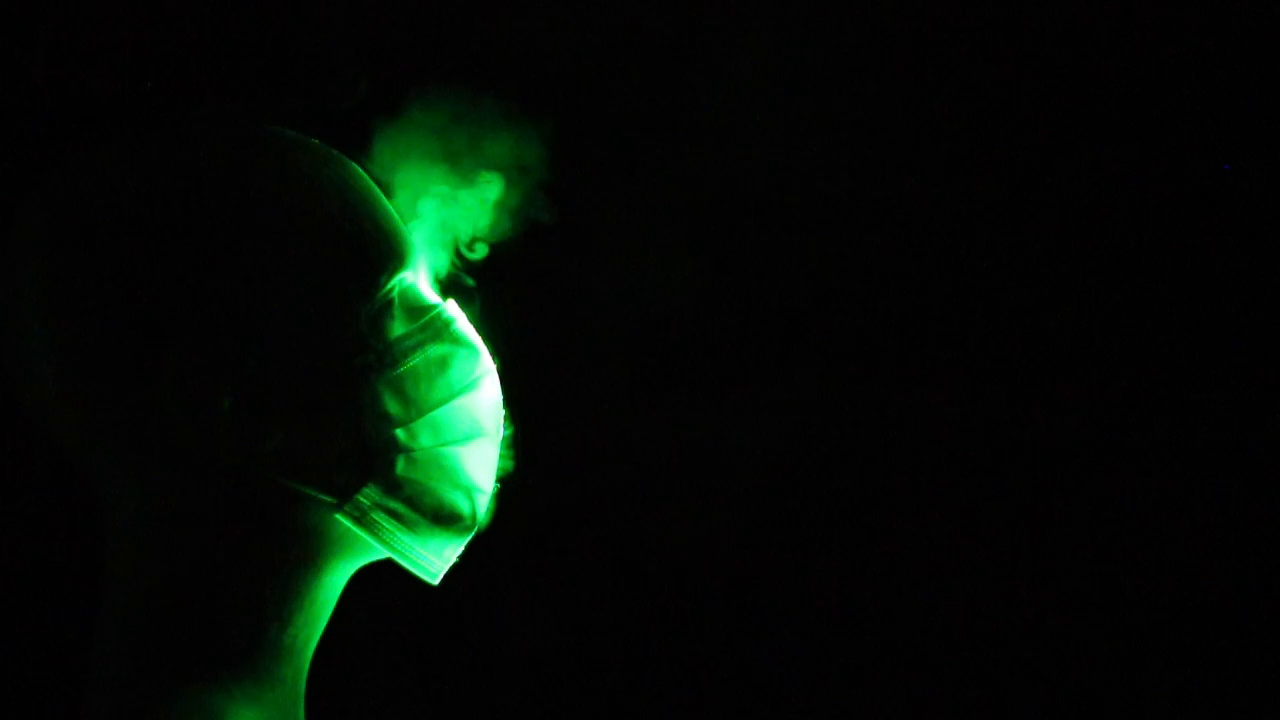}
\caption{\label{fig:surgA2}}
\end{subfigure}
\begin{subfigure}{0.49\linewidth}
\centering
\includegraphics[width=\linewidth]{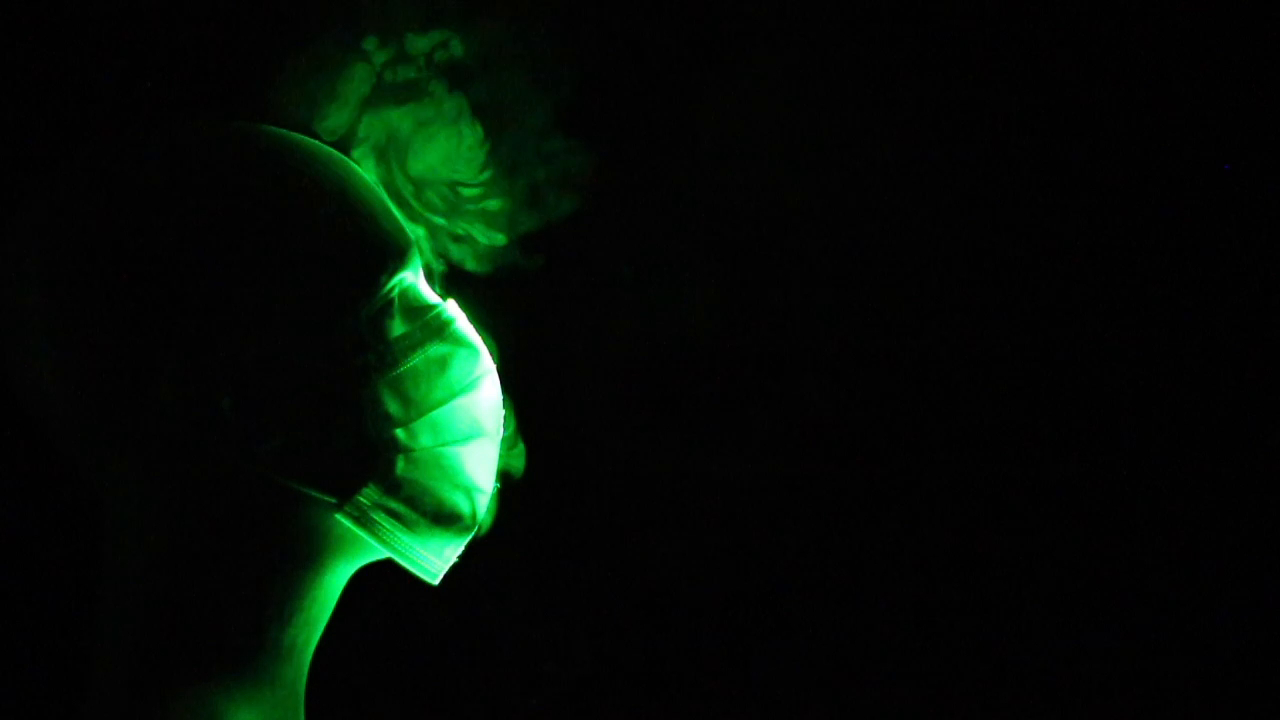}
\caption{\label{fig:surgA3}}
\end{subfigure}
\begin{subfigure}{0.49\linewidth}
\centering
\includegraphics[width=\linewidth]{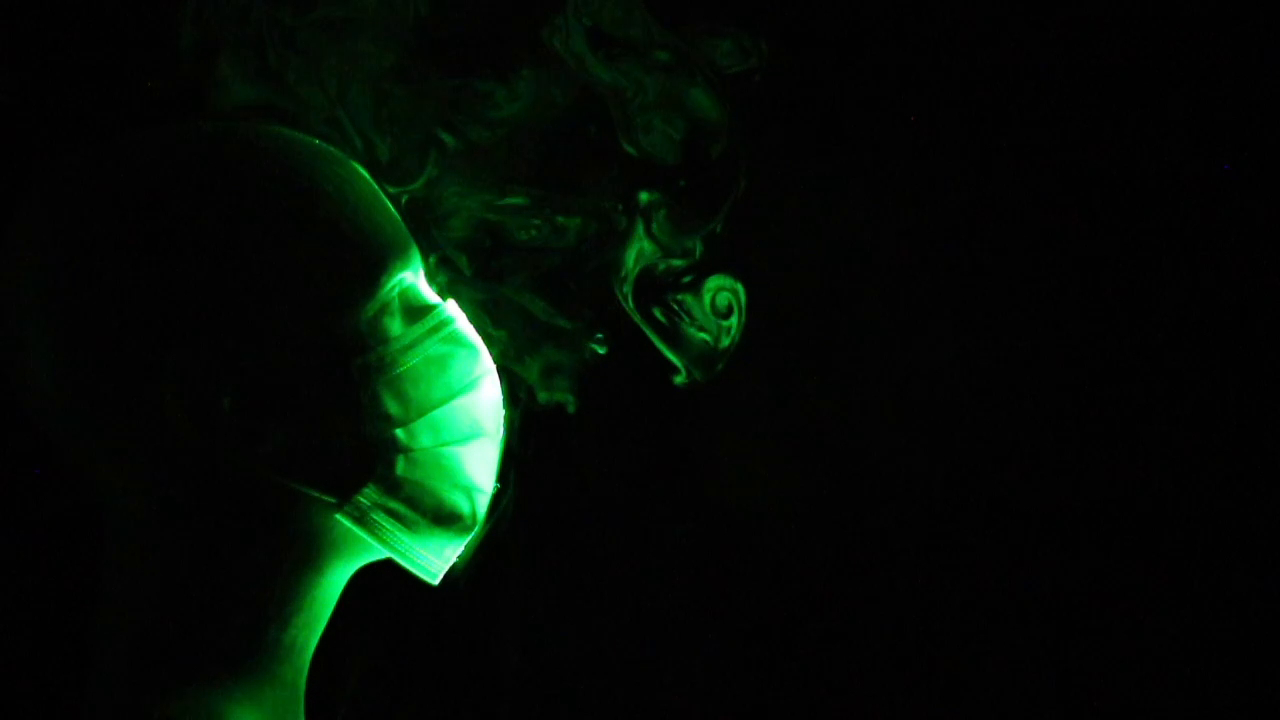}
\caption{\label{fig:surgA4}}
\end{subfigure}
\caption{\label{fig:surgA} Visualization of droplet \blue{spread} when a surgical mask (brand `A') is used to block the jet. (\subref{fig:surgA1}) Prior to emulating a cough/sneeze. (\subref{fig:surgA2}) 0.37 seconds after \blue{the} initiation of the emulated cough. (\subref{fig:surgA3}) 0.62 seconds. (\subref{fig:surgA4}) 2.33 seconds. (Multimedia View)}
\end{figure*}

\begin{figure*}[ht!]
\centering
\begin{subfigure}{0.49\linewidth}
\centering
\includegraphics[width=\linewidth]{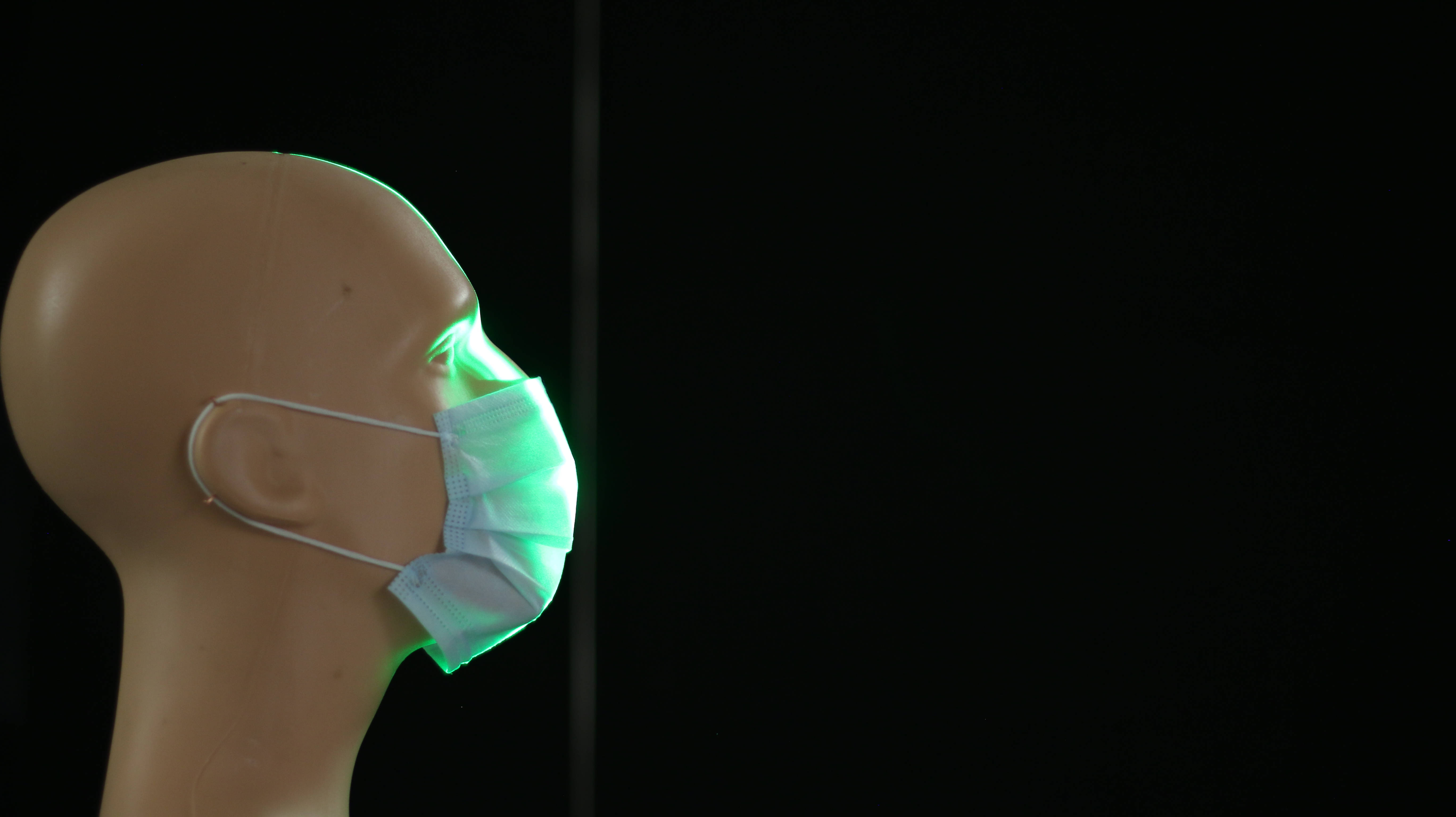}
\caption{\label{fig:surgB1}}
\end{subfigure}
\begin{subfigure}{0.49\linewidth}
\centering
\includegraphics[width=\linewidth]{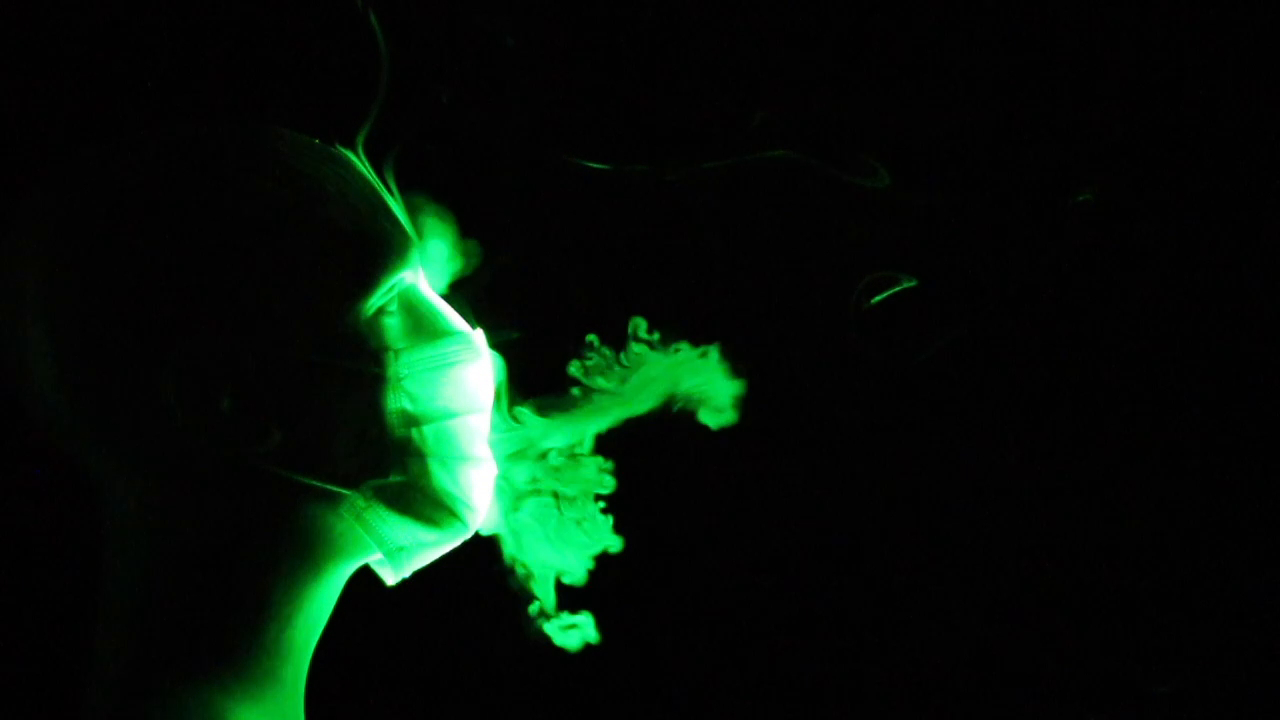}
\caption{\label{fig:surgB2}}
\end{subfigure}
\begin{subfigure}{0.49\linewidth}
\centering
\includegraphics[width=\linewidth]{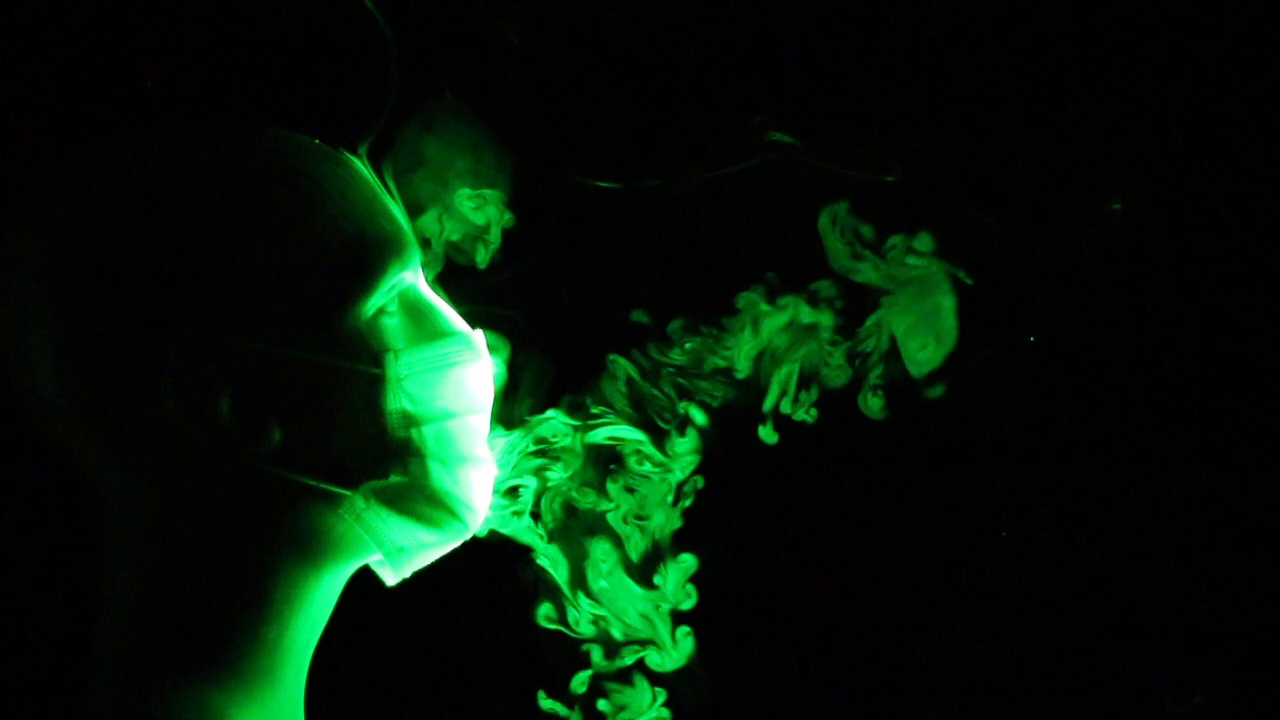}
\caption{\label{fig:surgB3}}
\end{subfigure}
\begin{subfigure}{0.49\linewidth}
\centering
\includegraphics[width=\linewidth]{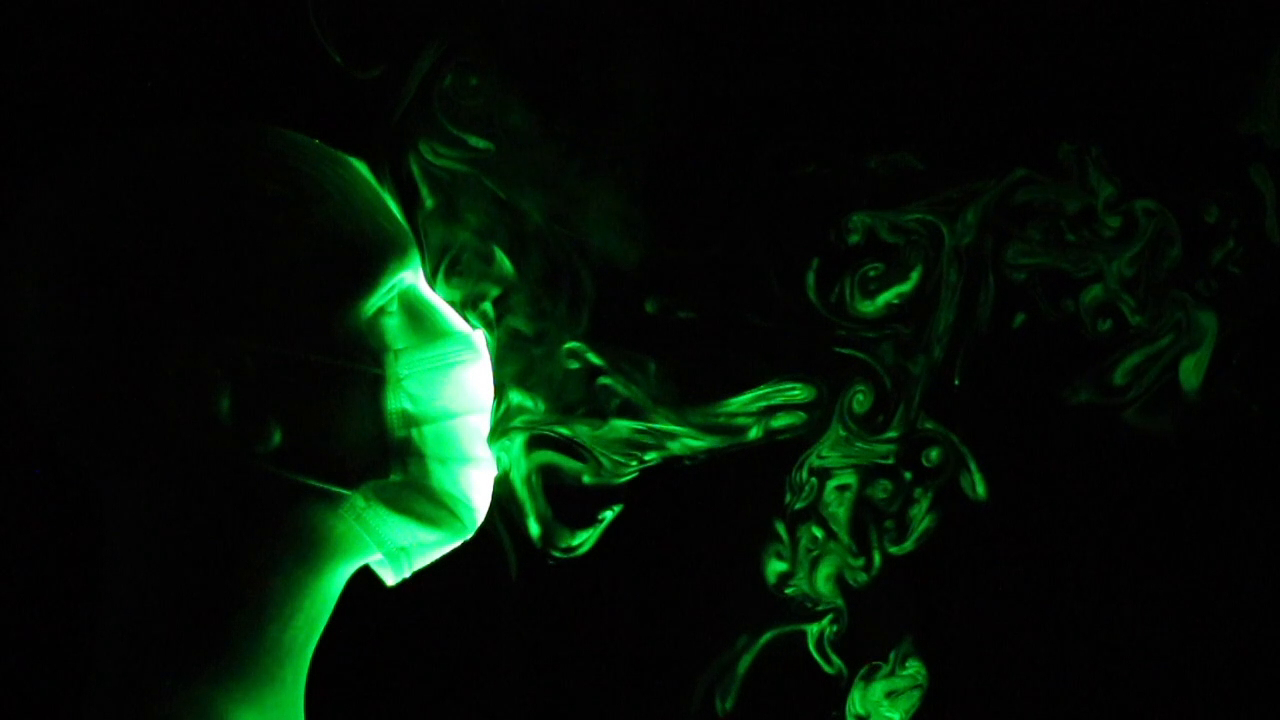}
\caption{\label{fig:surgB4}}
\end{subfigure}
\caption{\label{fig:surgB} Visualization of droplet \blue{spread} when a surgical mask (brand `B') is used to block the jet. (\subref{fig:surgB1}) Prior to emulating a cough/sneeze. (\subref{fig:surgB2}) 0.5 seconds after \blue{the} initiation of the emulated cough. (\subref{fig:surgB3}) 0.83 seconds. (\subref{fig:surgB4}) 3.13 seconds. (Multimedia View)}
\end{figure*}

In order to determine the performance of plain `surgical' masks in comparison to the N95-rated masks, we examine two different commercially available face masks in Figure~\ref{fig:surgA} (Multimedia View) and Figure~\ref{fig:surgB} (Multimedia View).
We note that neither of the two `surgical' masks tested here were recommended for medical use by the manufacturers. Such masks are becoming increasingly available commercially from a wide range of manufacturers, and they are seeing widespread adoption by the general population for regular use. We observe in Figure~\ref{fig:surgA} that the first surgical mask tested (brand `A') is very effective in stopping the forward progression of the jet. As expected, there is some leakage from the gap along the top of the mask, however it is not excessive, and it is comparable qualitatively to leakage from the regular N95-rated mask shown in Figure~\ref{fig:n95}. On the other hand, the second surgical mask (brand `B') which is shown in Figure~\ref{fig:surgB} displays significantly higher leakage of droplets through the mask material, and does not appear to be as effective as the first surgical mask (brand `A') in restricting droplet spread. This indicates that even among commercially available masks which may appear to be similar superficially, there can be significant underlying differences in the quality and type of materials used for manufacturing the masks.

To summarize, we have examined the effectiveness of face shields and masks equipped with exhalation ports in mitigating the spread of exhaled respiratory droplets. The aim of the qualitative visualizations presented here is to help increase public awareness regarding the effectiveness of these alternatives to regular masks. We observe that face shields are able to block the initial forward motion of the exhaled jet, however, aerosolized droplets expelled with the jet are able to move around the visor with relative ease. Over time, these droplets can disperse over a wide area in both the lateral and longitudinal \blue{directions}, albeit with decreasing droplet concentration. We have also compared droplet dispersal from a regular N95-rated face mask to one equipped with an exhale valve. As expected, the exhalation port significantly reduces the effectiveness of the mask as a means of source control, as a large number of droplets pass through the valve unfiltered. Notably, shields impede forward motion of the exhaled droplets to some extent, and masks with valves do so to an even lesser extent. However, once released into the environment, the aerosol-sized droplets get dispersed widely depending on \blue{light ambient disturbances}. Overall, the visuals presented here indicate that face shields and masks with exhale valves may not be as effective as regular face masks in restricting the spread of aerosolized droplets. Thus, despite the increased comfort that these alternatives offer, it may be preferable to use well-constructed plain masks. There is a possibility that widespread public adoption of the alternatives, in lieu of regular masks, could have an adverse effect on ongoing mitigation efforts against COVID-19.
\section*{Supplementary Material}
Please see supplementary material for additional videos regarding the effectiveness of various types of facemasks and a face shield.

\section*{Data Availability}
The data that supports the findings of this study is available within the article.

\bibliography{pof2020}

\end{document}